\documentclass[aps,twocolumn,showpacs]{revtex4}
\usepackage{graphicx}
\usepackage{epsfig}

\begin{document}
\title{Nonlinear dissipation can combat linear loss}

\author{D. Mogilevtsev$^{1,2}$, A. Mikhalychev$^2$, V. S. Shchesnovich$^1$ and N. Korolkova$^3$}

\affiliation{$^1$Centro de Ci\^encias Naturais e Humanas,
Universidade Federal do ABC, Santo Andr\'e,  SP, 09210-170 Brazil;
\\
 $^2$Institute of Physics, Belarus National Academy of Sciences,
F.Skarina Ave. 68, Minsk 220072 Belarus
\\
$^3$School of Physics and Astronomy, University of St Andrews,
North Haugh, St Andrews KY16 9SS, UK}

\begin{abstract}

We demonstrate that it is possible to compensate for effects of
strong linear loss when generating non-classical states by
engineered nonlinear dissipation. We show that it is always
possible to construct such a  loss-resistant dissipative gadget in
which, for a certain class of initial states, the desired
non-classical pure state can be attained within a particular time
interval with an arbitrary precision.  Further we demonstrate that
an arbitrarily large linear loss can still be compensated by a
sufficiently strong coherent or even thermal driving, thus
attaining a strongly non-classical (in particular, sub-Poissonian)
stationary mixed states.

\end{abstract}

\pacs{03.65.Yz, 42.50.Dv}

\maketitle

Nowadays, an engineered dissipation for quantum state
manipulation is an intensely developing field. More that a decade
ago, it has been shown that in the systems such as ions in magnetic traps
it is possible to tailor nonlinear dissipation in a
rather wide range \cite{zoll96}, and later the concept of the "quantum state protection"
was born \cite{dav2001}. Different kinds of a nonlinear
dissipative apparatas (aptly nicknamed "dissipative gadgets"
\cite{zoll2008,cirac2009}) have been shown to be useful for many
important tasks, for example, for generating non-classical and
entangled states of few-body and many-body systems
\cite{TFAb,TFComm,TFRepl,NlAbs,MS,plenio1999,kimble2000,parkins2003},
performing universal quantum computation
\cite{cirac2009,eisert2012}, constructing dissipatively protected
quantum memory \cite{cirac2011}, performing precisely timed
sequential operations, conditional measurements or error
correction \cite{eisert2012}. The central idea of all the dissipative
gadgets  is to make dissipation to drive the system towards a
desired steady state (which is practically independent of the
initial state). Remarkably, in this context, dissipation serves as a helpful quantum resource
rather than being a hindrance.

However, in all these currently known schemes,
the engineered dissipation is far from being an universal
efficient  tool for combating the usual linear loss, inevitably
present in any realistic dissipative gadget. For example, the
conventional single-photon loss makes the generation of a non-classical
pure stationary state impossible for just any kind of nonlinear dissipation
\cite{zoll2008}. Just as it is for the coherent control, one is
generally obliged to minimize linear losses using some extra effort while arranging for
the nonlinear terms to produce a desired effect (see, for example,
Ref. \cite{MS,zoll2012}). In particular, it is rather hard
to produce a desired state of an electromagnetic field
for schemes relying on optical nonlinearities
\cite{TFAb,TFComm,TFRepl,NlAbs,MS}. However, surprisingly enough, the nonlinear
dissipation, when properly designed, do can combat the effects of an arbitrarily strong linear
loss, for finite time intervals and for stationary states. That
is the main message of our contribution.

We start with demonstrating that it is always possible to
generate a state approximating the desired pure non-classical
state with any given precision for
an arbitrary ratio of linear and nonlinear loss rates. To
illustrate our argument, let us consider a simple Lindblad
master equation for the single mode of the electromagnetic field
\begin{equation}
\frac{d}{dt}\rho=-i[H,\rho]+\Gamma(\bar{n}+1)\mathcal{L}(a)\rho+
\Gamma\bar{n}\mathcal{L}(a^{\dagger})+\gamma\mathcal{L}(A)\rho
\label{master equation1}
\end{equation}
where the nonlinear dissipation  is described by the Lindblad operator
$A$, the operators $a^{\dagger},a$ are the usual bosonic creation
and annihilation operators, and  the operator $H$ represents the
system Hamiltonian. $\Gamma$ and $\gamma\geq 0$ are rates of
linear and  nonlinear dissipation, and  the parameter $\bar{n}$ represents
the average number of photons in the thermal pump. The Liouvillian $\mathcal{L}(x)$ acts on the density matrix as
$\mathcal{L}(x)\rho=2x\rho x^{\dagger}-x^{\dagger}x\rho-\rho
x^{\dagger}x$.

It has been proved that the coherent state is the only possible pure stationary state of Eq.(\ref{master equation1}) \cite{zoll2008}
for $\Gamma>0$ (and it will be the
vacuum state if no coherent driving is present). Still there is
a possibility to generate any desired state during the time interval
$t\ll 1/\Gamma$. To ilustrate   the principle, consider    $A=|\phi\rangle\langle y|a^k$, $k>1$, as  the Lindblad operator
for the  engineered dissipation, where the vector  $|y\rangle$ describes the initial state and $|\phi\rangle$ is the target  pure
state. The  density matrix satisfying  Eq.(\ref{master equation1}) with  the above nonlinear $A$-term has the property:
\begin{equation}
P\rho(t)P = \mathrm{Tr}\{P\rho(0)\}|0\rangle\langle0| + \mathcal{O}(e^{-\gamma_{\mathrm{eff}}t}).
\label{rho}
\end{equation}
Here  $\gamma_{\mathrm{eff}} =
\mathrm{min}\{1,2(1-|\langle\phi|\Psi\rangle|^2)\}\mathcal{N}\gamma$ with
$|\Psi\rangle=(\mathcal{N})^{-1/2}(a^{\dagger})^k|y\rangle$. The norm-factor is given by $\mathcal{N}=\langle
y|a^k(a^{\dagger})^k|y\rangle$. The state $|0\rangle =
(1-|\langle\phi|\Psi\rangle|^2)^{-1/2}(|\phi\rangle
-\langle\Psi|\phi\rangle|\Psi\rangle)$ is the orthogonal
complement of $|\Psi\rangle$  in the subspace spanned by
$|\phi\rangle$ and $|\Psi\rangle$, with $P$ being the projector on
this subspace.  In particular, Eq.(\ref{rho}) gives
\begin{equation}
\rho_{\phi \phi}(\infty) = (1-|\langle\phi|\Psi\rangle|^2) \mathrm{Tr}\{P\rho(0)\},
\label{steady1}
\end{equation}
that is, for $P\rho(0)=\rho(0)$  the large $\mathcal{N}$  in
Eq.(\ref{rho})  implies that the system will rapidly evolve  to
the state $|\phi\rangle$ with  the fidelity
$1-|\langle\phi|\Psi\rangle|^2$. Furthermore, note that if
$\langle\phi|\Psi\rangle = 0$, one can  drive the system from an initial
state $|\Psi\rangle$ to the state $|\phi\rangle$ orthogonal to it.

\begin{figure}[tbh]
\includegraphics[width=9.5cm]{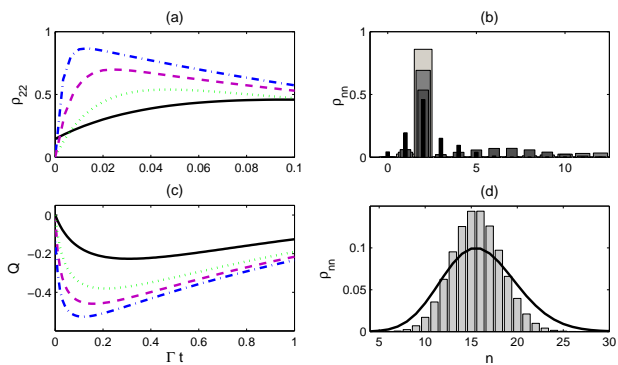}
\caption{(Color online) Generation of nonclassical states from a
coherent state $\vert \alpha \rangle$ in the presence of linear
loss using designed nonlinear dissipation described by the
Lindblad operator $A$. Thermal driving is absent, ${\bar n}=0$,
and the loss ratio is given by $\gamma=\Gamma/5$. I. Generation of
the Fock state $\vert 2 \rangle$ using nonlinear dissipation given
by $A=|2\rangle\langle \alpha|a^2$. (a) Fidelity of the target
two-photon state as a function of nonlinear loss $\Gamma t$.
Solid, dotted, dashed and dash-dotted lines correspond to
$\alpha=2,3,4,5$. The panel (b) shows the photon number
distribution of the generated state when the fidelity in (a) is
maximal; black, dark grey, grey and light bars correspond to
$\alpha=2,3,4,5$. II. Generation of sub-Poissonian light and
photon-number squeezing using tailored dissipation with
$A=a(a^+a-1)$. (c) Mandel Q-parameter vs nonlinear loss $\Gamma
t$; solid, dotted, dashed and dash-dotted lines correspond to
$\alpha=2,4,6,8$. (d) Photon number distribution of the generated
state for the minimal Q and $\alpha = 8$ (grey bars) in comparison
with the Poissonian distribution with the same average number of
photons (solid line).  }\label{fig1}\end{figure} For a linear loss
present,  a simple physical picture behind the loss-suppression
mechanism is provided by system jumps to the lower energy levels.
Notably, the rate of transition to the lower levels due to
nonlinear dissipation can be much higher than the transition rate
due to linear loss. Now, the stationary state of the evolution
induced by nonlinear dissipation corresponds to a nonclassical
state. Hence when the dynamics due to nonlinear dissipation
prevails over that due to linear dissipation, the system is driven
into a desired nonclassical state with influence of linear loss
being negligible. Designing dissipation (i.e., the Lindblad
operator) we determine the form of the nonclassical target state.
If we apply this quantum jump formalism to the system desribed by
Eqs.~(\ref{master equation1})-(\ref{steady1}), we can deduce that
for
\begin{equation}
\Gamma(\bar{n}+1)\langle
\Psi|a^{\dagger}a|\Psi\rangle\ll\gamma\langle\Psi|A^{\dagger}A|\Psi\rangle
\label{condition}
\end{equation}
the influence of the linear loss (and of incoherent driving)
during the time of  the target state generation is negligible.
For a
coherent initial state $|y\rangle = |\alpha\rangle$ with sufficiently large
amplitude, $\alpha\gg1$, condition (\ref{condition}) gives
$\Gamma(\bar{n}+1)\ll\gamma\mathcal{N}$. In practical terms, for
the target state lying within the subspace of Fock vectors with up
to $k$ photons, the generation can be nearly perfect and improved
further just by increasing the amplitude $\alpha$.

Fig.~\ref{fig1}~(a,b) illustrates the solution of Eq.(\ref{master
equation1}) for the generation of the two-photon Fock state
from an initial coherent state $|\alpha\rangle$ with the nonlinear dissipation described by the
Lindblad operator  $A=|2\rangle\langle \alpha|a^2$.
The rate of nonlinear loss, $\Gamma$, is $5$ times
less than the rate of the linear loss, $\gamma$. The incoherent driving is
absent. By increasing the
amplitude $\alpha$ the target state $|2\rangle$ is approximated with a larger fidelity  and
over a shorter time period, despite the presence of rather strong
linear loss (Fig.\ref{fig1}(a,b)).

As we have seen, if for a class of states the jump rate of the
nonlinear dissipation far exceeds the jump rate for the linear
loss, then an influence of the linear loss is negligible.
Moreover,  the dynamics due to the engineered nonlinear
dissipation can also be non-exponentially fast. For example, for
the two-photon dissipation, quite different initial states can
decay into the stationary state practically over the same period
of time \cite{DodMiz}. Thus, for an appropriately constructed
dissipative gadget it is always possible to combat the linear loss
with any pre-defined rate just by choosing the appropriate initial
state, in particular, by increasing an average number of photons
in the initial state. Of course, there are diffirent practical
limitations in construction of dissipative gadgets. However, even
for available types of nonlinear dissipative processes  it is
possible to have significant non-classicality (in particular,
large photon-number squeezing) on the time scales when influence
of the linear losses is negligible (see, for example, Refs.
\cite{MS,valera2011}). This   is illustrated in Fig.\ref{fig1}(c,
d), where  the exact solution of Eq.(\ref{master equation1}) is
shown for the nonlinear dissipation described by the Lindblad
operator $A=a(a^+a-1)$ in  the presence of a linear loss and for
an initial coherent state $|\alpha\rangle$. Obviously, this
Lindblad operator   satisfies   condition (\ref{condition}) for a
sufficiently large $\alpha$. In the initial stages of  evolution,
nonclassical states can be generated  and the non-classicality
(photon-number squeezing in this case) is increased with
increasing of the amplitude $\alpha$ in the initial state. Panel
(c) shows the Mandel Q parameter confirming that the generated
light exhibit the sub-Poissonian statistics (for $Q< 0$). Note,
that the maximum of non-classicality is reached rather quickly.
The generated state then further evolves towards the classical
regime under the influence of the linear loss but it happens quite
slowly compared to the time required for the nonclassical state
generation. This is a rather general feature of the dynamics for
any dissipative gadget satisfying  condition (\ref{condition})
\cite{mikhalychev new}.

\begin{figure}[tbh]
\includegraphics[width=9.5cm]{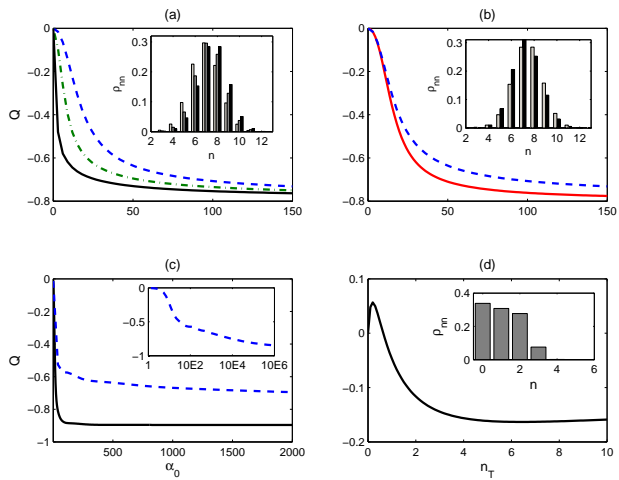}
\caption{(Color online) Generation of nonclassical states  in the long-time limit using nonlinear coherent loss (NCL)
to counteract linear loss.
(a) Mandel's Q-parameter vs amplitude of the coherent driving $\alpha_0=\Omega/\gamma$ for
the NCL with $A=aa^{\dagger}a-1$ provided by the exact solution
of Eq.(\ref{master equation1}); solid, dash-dotted and dashed
curves correspond to the relative loss rate
$\Gamma=\gamma$, $5\gamma$, $10\gamma$. Inset shows photon-number
distribution of the stationary state for $\alpha_0=150$;
light-grey, grey and black bars correspond to
$\Gamma=\gamma$, $5\gamma$, $10\gamma$. (b) Same for the NCL with $A=a(a^{\dagger}a-1)$ for the exact
solution of Eq.(\ref{master equation1}) (dashed line) and for
the solution of the approximate equation (\ref{master
equation2})(solid line). Inset shows photon-number distributions
for the solution of the exact (grey bars) and of the
approximate (black bars) equations. (c) Exact (dashed line) and
approximate (solid line) solutions for Q-parameter for
$A=a(a^{\dagger}a-1)^2$; the inset shows long-time behavior of
Q-parameter given by the exact solution. (d) Exact solution for
Q-factor in dependence of the average number of thermal pumping
photons, $n_T$; the coherent driving is absent;
$A=a(a^{\dagger}a-1)^3$. }\label{fig2}\end{figure}

It might seem surprising, but analogous approach
can allow for combating linear losses also in
the long-time limit. More precisely, there is a class of nonlinear
losses such that for an arbitrarily large (but finite) ratio
$\varepsilon=\Gamma/\gamma$ an influence of losses on the
stationary state can be completely eliminated by the sufficiently
strong coherent driving, and, moreover, even by thermal pumping. Note,
that the steady state, while being mixed, can still be strongly
non-classical. Consider, for the example, nonlinear
coherent loss (NCL) described by the Lindblad operators
$A=af(a^{\dagger}a)$, $f(x)$ being a smooth function. This
operator is the annihilation operator for the so called
"f-deformed" quantum harmonic oscillator; eigenstates of this
operator are reffered to as "nonlinear coherent states" \cite{manko}.
Any pure state non-orthogonal to an arbitrary Fock states can be
exactly represented as a nonlinear coherent state. If it is
orthogonal to some Fock states, then one can still devise a
nonlinear coherent state closely approximating the state in
question \cite{kis}. NCL can be realized in practice in a number
of schemes (for example, with ions or atoms in traps
\cite{zoll96}; Bose-Einstein condensates \cite{SM} or even in
multicore nonlinear optical fibres \cite{MS}. With function
$f(x)$ having a finite or countable number of zeros, one can
generate Fock states or even "comb" the initial state by filtering
out some pre-defined set of components in the given basis
\cite{michalychev2011}.

Let us show now that even a weak NCL can be protected
from large linear loss and used for generation of non-classical
stationary states. We use the classical driving in the
form $H=i\Omega(a-a^{\dagger})$ where the parameter $\Omega$
represents the strength of driving and, for simplicity, is taken
to be real. For the moment being, the thermal pumping is
assumed to be absent, $\bar{n}=0$. Fig.~\ref{fig2} depicts such a
non-classicality rescue procedure when generating sub-Poissionian light.
Notably, for $A=a(a^{\dagger}a-1)$, the
Mandel's Q parameter $Q=(\langle (a^{\dagger}a))^2\rangle-\langle
(a^{\dagger}a))\rangle^2)/\langle (a^{\dagger}a))\rangle-1$
eventually tends to the same limiting value for quite different
linear loss rates, $\Gamma=\gamma$, $5\gamma$ and $10\gamma$ (Fig.~\ref{fig2}(a)).  It is remarkable that
the generated states are also quite similar (see inset in
Fig.\ref{fig2}(a) for the density matrix).

The nature of this phenomenon can be well illustrated and
clarified with the help of the following simple approximation.
Eq.(\ref{master equation1}) can be written as
\begin{eqnarray}
\nonumber \frac{d}{dt}\rho= \gamma
a\rho(B^{\dagger}-\alpha_0)+\gamma(B-\alpha_0)\rho a^{\dagger}-\\
\gamma a^{\dagger}(B-\alpha_0)\rho-\gamma
\rho(B^{\dagger}-\alpha_0)a -\\
\nonumber
\gamma a [[\rho, f(a^\dagger a)],f(a^\dagger a)]
a^\dagger
 \label{master equation2}
\end{eqnarray}
where $B=a([f(a^{\dagger}a)]^2+\varepsilon)$.  The last term is
small in comparison with the others when the inequality
$\{f(n)-f(m)\}^2 \ll 2 f(n)f(m)$ holds for all $n$ and $m$,
corresponding to essentially non-zero density matrix elements
$\rho_{nm}$. For example, for any power-law nonlinearity $f(x)\sim
x^k$, this condition is satisfied when the driving is sufficiently
strong. Neglecting the last term in Eq.(\ref{master equation2}),
the stationary state, $\rho^s$, for Eq.(\ref{master equation2}) is
the eigenstate of the operator $B$ satisfying
$B\rho^s=\alpha_0\rho^s$ and $\rho^s B^{\dagger}=\alpha_0\rho^s$;
where $\alpha_0=\Omega/\gamma$. This leads to the following
recurrence relation for the diagonal (in the Fock-state basis)
elements of the steady state:
\begin{equation}
\rho_{n,n}^s=\rho_{n-1,n-1}^s
\frac{\alpha_0^2}{n(f(n)^2+\varepsilon)^2}. \label{recurrence}
\end{equation}
Assuming that the steady-state photon number distribution is
peaked at $n_0$, from Eq.(\ref{recurrence}) we get a simple
condition for determining $n_0$:
\[\rho_{n,n}^s\approx\rho_{n-1,n-1}^s\Rightarrow n_0((f(n_0)^2+\varepsilon)^2)\approx\alpha_0^2.\]

This condition indicates that for sufficiently strong classical
driving and for function $f(n)$ increasing monotonically for
sufficiently large $n>0$, one will always have
$f(n_0)^2\gg\varepsilon$, and the influence of linear losses on
the generated steady state will be negligible. The relation
(\ref{recurrence}) also allows for estimating the width of the
photon-number distribution of the stationary state. Assuming that
the value of the diagonal element $\rho_{n,n}^s$ changes only weakly
for small changes $\delta n$ of the photon number $n$ near
$n_0$, one obtains from Eq.(\ref{recurrence}) the following
expression:
\begin{eqnarray}
\rho_{n_0+\delta n,n_0+\delta n}^s\approx
\rho_{n_0,n_0}^s\times \nonumber
\\
\exp\left\{ -\frac{|\delta n|(|\delta n|+1)}{2n_0}\left(
1+\frac{4n_0\dot{f}(n_0)f(n_0)}{f(n_0)^2+\varepsilon}  \right)
\right\}. \label{estimate}
\end{eqnarray}
From Eq.(\ref{estimate}) we can estimate the variance,
$\Delta^2n$, of the steady-state photon number distribution:
\begin{equation}
\Delta^2n\approx n_0\left(
1+\frac{4n_0\dot{f}(n_0)f(n_0)}{f(n_0)^2+\varepsilon}
\right)^{-1}. \label{variance}
\end{equation}

Thus, our approximation shows that in the limit of strong driving
the steady state produced by the NCL will always be photon-number
squeezed state, provided that $\dot{f}(n_0)>0$. The squeezing can be quite
high. For example, let us use
$f(n)=n-1$, which has been proved feasible using three-well potential in
Bose-Einstein condensates \cite{SM} or three-core nonlinear fibers
\cite{MS}. For the simple NCL with $f(n)=n-1$, the value
$\Delta^2n\rightarrow n_0/5$ is asymptotically reached for $n_0\approx
(\alpha_0)^{2/5}\rightarrow+\infty$. This value corresponds to the Mandel
parameter equal to $-0.8$. Fig.~\ref{fig2}(b) shows that
the approximation to Eq.~(\ref{master
equation2}) indeed gives rather good estimate for both the Mandel
parameter and the generated state. It should be noted though, that for
rapidly increasing $f(n)$ the approximation works somewhat worse. It
can be easily seen from Eq.(\ref{variance}), since it predicts
Q-parameter close to $-1$ for rapidly increasing $f(n)$, e.g.,
$Q\sim -4k/(4k+1)$ for $f(n)=n^k$. $Q=-1$ corresponds to
the Fock state. But when the photon
number distribution becomes very narrow, the assumption of slowly changing $\rho_{nn}$
near maximum of the photon number distribution can be hardly
applied. Nevertheless, the approximation still provides
a qualitatively correct description as illustrated in Fig.~\ref{fig2}(c) for
$f=(x-1)^2$, where exact and approximate solutions for Q-parameter
are compared. Hence even for $f(x)$ increasing rather rapidly one can indeed
have highly pronounced non-classicality unaffected by any linear
loss. However, it happens at the expense of quite strong coherent driving required
(see Fig.~\ref{fig2}(c)).

The essence of our method to preserve non-classicality lies in engineering the evolution
to yield a stationary state obeying the following requirement.
A stationary state of the master equation for a given system should be such, that the
transition rates corresponding to nonlinear loss far exceed
transition terms corresponding to linear loss, that is, the condition
(\ref{condition}) holds. This simple fact points to
the possibility of exploiting not only the coherent driving, but other
kinds of driving, too. In particular, the conventional linear
thermal driving can protect non-classicality as well,
although this seems quite paradoxical.
Indeed, in the absence of the coherent
driving ($\Omega=0$), from exact Eq.(\ref{master
equation2}) the following recurrence relation is obtained:
\begin{equation}
\label{recurrence1} \rho_{n}=\rho_{n-1} \frac{\bar n} {(\bar n+1)
+ (\gamma/\Gamma) f^2(n) }
\end{equation}
For small values of $(\gamma/\Gamma) f^2(n)$ the state (\ref{recurrence1}) is very close to a thermal state.
For $(\gamma/\Gamma) f^2(n) \gg (\bar n+1)$ the photon number
distribution is effectively truncated. The truncation number, $n_t$
can be estimated from the following equation: $\bar n + 1 \sim
(\gamma / \Gamma) f(n_t)$. If the density matrix elements $\rho_n$
decrease with increasing photon number $n$ faster than for any
coherent state: $\rho_n / \rho_{n-1} = o(1/n)$, the considered
state is nonclassical. Eq.(\ref{recurrence1}) implies that for
$f(x)$ growing faster than $x$ this condition is satisfied and the
stationary state is nonclassical for any values of
system parameters. Fig.~\ref{fig2}(d) shows an example of such a
"thermal rescue"  for the nonlinear loss with
$f(x)=(x-1)^3$. Thus, remarkably, the thermal excitation is able to
produce photon anti-bunching. However, it is easy to see that the
minimal $Q$-value always corresponds to some finite $\bar n$.

To conclude, we have shown that dissipative gadgets can be made
extremely robust with respect to additional linear losses
unavoidably affecting any realistic engineered dissipation scheme.
First we have shown that
it is always possible to devise a dissipative gadget
for the
generation of any desired target state starting from the wide class of initial
states during the time interval when influence of linear loss is negligible.
We provided several examples illustrating such scenario using coherent input states.
The larger the
difference in average number of photons between the initial state
and the target state, the more robust the scheme can be made.
Then, by applying conventional linear quasiclassical driving,
the non-classicality of the stationary state
generated by the dissipative gadget can be preserved. It is remarkable that there exists
a class of dissipative gadgets based on the nonlinear
coherent loss (NCL), for which the influence of linear loss can be
completely annihilated by a sufficiently strong coherent
driving. The resulted state is mixed. However, this stationary mixed state can be very
close to the Fock state exhibiting strong photon-number squeezing.
Furthermore, there exists a class of dissipative gadgets, for which
non-classicality of the stationary state can be rescued even by using
usual incoherent thermal driving.

This work was supported by the Foundation of Basic Research of the
Republic of Belarus, by the National Academy of Sciences of
Belarus through the Program "Convergence", by the Brazilian
Agency FAPESP (project 2011/19696-0) (D.M.) and has received funding
from the European Community's Seventh Framework
Programme (FP7/2007-2013) under grant agreement n$^\circ$
270843 (iQIT).

\end{document}